\documentclass[twocolumn,showpacs,preprintnumbers,amsmath,amssymb]{revtex4}

\usepackage{wasysym}
\usepackage{graphicx}
\usepackage{dcolumn}
\usepackage{bm}

\begin{document}

\title{Dynamical properties of a trapped dipolar Fermi gas at finite temperature}
\author{J.-N. Zhang$^1$, R.-Z. Qiu$^1$, L. He$^2$, and S. Yi$^1$}

\affiliation{$^1$Key Laboratory of Frontiers in Theoretical Physics, Institute of Theoretical Physics, Chinese Academy of Sciences, Beijing 100190, China}
\affiliation{$^2$Institut f\"{u}r Theoretische Physik, Johann Wolfgang Goethe-Universit\"{a}t, 60438 Frankfurt/Main, Germany}

\begin{abstract}
We investigate the dynamical properties of a trapped finite-temperature normal Fermi gas with dipole-dipole interaction. For the free expansion dynamics, we show that the expanded gas always becomes stretched along the direction of the dipole moment. In addition, we present the temperature and interaction dependences of the asymptotical aspect ratio. We further study the collapse dynamics of the system by suddenly increasing the dipolar interaction strength. We show that, in contrast to the anisotropic collapse of a dipolar Bose-Einstein condensate, a dipolar Fermi gas always collapses isotropically when the system becomes globally unstable. We also explore the interaction and temperature dependences for the frequencies of the low-lying collective excitations.
\end{abstract}

\date{\today}

\maketitle

\section{Introduction}
The experimental progress towards making quantum degenerate gases of fermionic polar molecules~\cite{ye1, ye2, dwang} have stimulated many interests in theoretical studying of the dipolar Fermi gases. Compared to short-range $s$-wave interaction, the long-range and anisotropic nature of the dipole-dipole interaction may give rise to new phenomena in degenerate Fermi gases, such as the anisotropic Bardeen-Cooper-Schrieffer pairing~\cite{you,bara,bara2,pu,samo,wu,shi} and the strongly correlated quantum phases in rapidly rotating traps~\cite{bara3,oste} and in optical lattices~\cite{liu,quin,he2}.

For dipolar Fermi gases in normal phase, many theoretical studies focus on the zero-temperature regime. In early stage, the momentum distribution was assumed to be isotropic~\cite{goral2,goral3,he}. As a result, the Fock-exchange interaction was completely ignored. Miyakawa {\it et al}.~\cite{miya} introduced an ellipsoidal ansatz for the phase-space distribution function (PSDF) which allows for the deformation in the momentum distribution. It was then found that Fock-exchange interaction induces a momentum-space deformation, which has an important impact on the stability of the system. The variational approach can be easily implemented, but it often overestimates the stability of the system as it is unable to capture the local collapses. Two of us then solve the semiclassical theory by numerically minimizing the total energy of the system~\cite{zhang1}. In addition, Chan {\it et al}.~\cite{chan} have developed an analytical Fermi liquid description for the dipolar Fermi gas. It was also predicated that this system may support biaxial nematic phases~\cite{freg,freg2} and an interaction-driven quantum phase transition from a paraelectric to a ferroelectric quantum gas~\cite{lin}.

As to the dynamical properties of the zero-temperature dipolar Fermi gases, Sogo {\it et al}.~\cite{sogo} have investigated the collective excitations and free expansions based on the ellipsoidal ansatz for the equilibrium distribution function. Lima and Pelster~\cite{lima,lima2} have presented a detailed theory on the variational time-dependent Hartree-Fock approach, which was subsequently used to study the static and dynamical properties of a dipolar Fermi gas in the hydrodynamic regime. Ronen and Bohn~\cite{bohn} have also investigated the zero sound propagation in a homogeneous dipolar Fermi gas.

For finite temperature systems, two of us have studied the properties of an equilibrium dipolar gas through a full numerical calculation~\cite{zhang2}. Baillie and Blakie have investigated the first- and second-order correlation properties~\cite{bail}. In high-temperature (non-degenerate) regime, Endo {\it et al}.~\cite{endo} have also introduced a variational ansatz which can describe the deformations in both real- and moment-space. So far, only the equilibrium state properties have been explored in the finite-temperature regime.

Experimentally, the detections of the dipolar effects in Bose-Einstein condensates often rely on their dynamical behaviors, for instance, the free expansion~\cite{pfau} and the low-lying collective excitations~\cite{bism}. For zero-temperature dipolar Fermi gases, the dynamical properties were theoretically investigated in Ref.~\cite{sogo,lima} based on the variational calculations. In this paper, we study the dynamical properties of a finite-temperature dipolar Fermi gas, to extend the works in Refs.~\cite{sogo,zhang2}. Using semiclassical theory, our system is described by the PSDF, whose dynamical behavior is governed by the Boltzmann-Vlasov equation. In order to make the numerical calculation manageable while still maintain the necessary accuracy, we will employ a hybrid approach. Namely, the equilibrium distribution function is obtained via a full numerical calculation; while the dynamical equation is solved by making use of the scaling ansatz. Our particular interests are focused on three types of the dynamical behaviors: free expansion, collapse dynamics, and low-lying collective excitations. We will show how these dynamical properties depend on temperature and dipolar interaction.

The content of this paper is organized as follows. In Sec.~\ref{equil}, we present our model Hamiltonian for a trapped single-component dipolar Fermi gas and outline the semi-classical description for the equilibrium state of the system at finite temperature. In Sec.~\ref{bvescal}, we introduce the Boltzmann-Vlasov equation and its scaling solution. We then derive a set of equations for the scaling parameters which describe dynamical behaviors of the system. The numerical results on free expansion, collapse dynamics, and low-lying collective excitations are presented in Sec.~\ref{resu}. Finally, we conclude in Sec.~\ref{concl}.

\section{Phase-space distribution function of an equilibrium state}\label{equil}
We consider a system of $N$ single-component fermionic polar molecules trapped in a harmonic potential
\begin{eqnarray*}
U_{\rm ho}\left({\mathbf r}\right)=\frac{1}{2}M\left(\omega_x^2x^2 +\omega_y^2y^2+\omega_z^2z^2\right),
\end{eqnarray*}
where $M$ is the mass of the molecule and $\omega_\eta$ ($\eta=x,y,z$) are the trap frequencies. For simplicity, the trapping potential is assumed to be axially symmetric with $\omega_x=\omega_y=\omega_\perp$. The shape of the trap is then characterized by the aspect ratio $\lambda=\omega_z/\omega_\perp$. We assume that each molecule possesses a permanent electric dipole moment $d$ which is polarized along the positive $z$-axis by an external electric field, such that the inter-particle dipolar interaction potential becomes
\begin{eqnarray}
V_d\left({\mathbf r}\right)=c_d\frac{x^2+y^2-2z^2}{(x^2+y^2+z^2)^{5/2}},
\end{eqnarray}
where the dipolar interaction strength is characterized by $c_d=d^2/\left(4\pi\varepsilon_0\right)$ with $\varepsilon_0$ being the permittivity of free space. Since the $s$-wave scattering length vanishes for spin polarized fermions, the Hamiltonian for the system under consideration takes the form
\begin{eqnarray}
\hat H=\sum_{i=1}^N\left[-\frac{\hbar^2\nabla_i^2}{2M}+U_{\rm ho}\left({\mathbf r}_i\right)\right]+\frac{1}{2}\sum_{i\neq j=1}^NV_d\left({\mathbf r}_i-{\mathbf r}_j\right).\label{eq:hamil}
\end{eqnarray}

Within the framework of semiclassical theory, the system is described by the PSDF $f\left({\mathbf r},{\mathbf k}\right)$, which, for an equilibrium state, satisfies the Fermi-Dirac statistics
\begin{eqnarray}
f\left({\mathbf r},{\mathbf k}\right)=\left[{\rm exp}\left(\frac{\varepsilon\left({\mathbf r},{\mathbf k}\right)-\mu}{k_BT}\right)+1\right]^{-1},\label{eq:Fermi_Dirac}
\end{eqnarray}
where $k_B$ is the Boltzmann constant, $T$ is the temperature, and $\mu$ is the chemical potential introduced to conserve the total number of particles
\begin{eqnarray}
N=(2\pi)^{-3}\int d{\mathbf r}d{\mathbf k}f(\mathbf{r},\mathbf{k}),\label{eq:normal}
\end{eqnarray}
and 
\begin{eqnarray}
\varepsilon({\mathbf r},{\mathbf k})=\frac{\hbar^2{\mathbf k}^2}{2M}+U_{\rm eff}({\mathbf r},{\mathbf k})\label{quasipot}
\end{eqnarray}
is the quasi-particle dispersion relation. The effective potential $U_{\rm eff}$ contains the contributions from the external trapping potential and the mean field induced by the inter-particle dipolar interaction
\begin{eqnarray}
U_{\rm eff}\left(\mathbf{r},\mathbf{k}\right)&=&U_{\mathrm{ho}}\left( \mathbf{
r}\right)+\int d{\mathbf r}V_d\left( {\mathbf{r}}-{\mathbf{r}}
^{\prime }\right) n({\mathbf r}')\nonumber\\
&&-\int \!\frac{d{\mathbf k}'}{(2\pi)^{3}}
\widetilde V_d\left({\mathbf k}-{\mathbf k}'\right)f\left(\mathbf{r},\mathbf{k}^{\prime }\right),\label{eq:ueff}
\end{eqnarray}
where $n\left({\mathbf r}\right)=\left(2\pi\right)^{-3}\int d{\mathbf k}f\left({\mathbf r},{\mathbf k}\right)$ is the real-space density and $$\widetilde V_d\left({\mathbf k}\right)=-c_d\frac{4\pi}{3} \frac{k_x^2+k_y^2-2k_z^2}{k_x^2+k_y^2+k_z^2}$$ is the Fourier transform of $V_{d}({\mathbf r})$. The second and third terms on the right-hand-side of Eq.~(\ref{eq:ueff}) originate, respectively, from the Hartree-direct and Fock-exchange interactions.

Equations (\ref{eq:Fermi_Dirac})-(\ref{eq:ueff}) form a closed system of equations which can be solved numerically through an iterative procedure to obtain an equilibrium PSDF. 

\section{Boltzmann-Vlasov equation and scaling ansatz}\label{bvescal}
The dynamical behavior of the system is described by the time-dependent PSDF $f({\mathbf r},{\mathbf k},t)$, which, in the collisionless regime, satisfies the Boltzmann-Vlasov kinetic equation~\cite{sogo,stringari1}
\begin{eqnarray}
\frac{\partial f\left( \mathbf{r},\mathbf{k},t\right) }{\partial t}&+&
\left(\frac{\hbar
\mathbf{k}}{M}+\frac{1}{\hbar}\frac{\partial}{\partial
\mathbf{k}}U_{\rm eff}\left( \mathbf{r},\mathbf{k},t\right)\right)
\cdot \frac{\partial }{\partial \mathbf{r}}f\left( \mathbf{r},
\mathbf{k},t\right)\nonumber\\
&-&\frac{1}{\hbar}\frac{\partial }{\partial \mathbf{r}}U_{\rm eff}\left( \mathbf{r},\mathbf{k} ,t\right) \cdot \frac{\partial }{\partial \mathbf{k}}f\left( \mathbf{r},
\mathbf{k},t\right)=0.\label{bve}
\end{eqnarray}
Here, the effective potential $U_{\rm eff}({\mathbf r},{\mathbf k},t)$ is also determined by Eq.~(\ref{eq:ueff}) with $n({\mathbf r})$ and $f({\mathbf r},{\mathbf k})$ being replaced by the time-dependent ones. 

Directly evolving the Boltzmann-Vlasov equation in phase space requires tremendous numerical efforts. To simplify the calculation, we make use of the scaling ansatz which assumes that Eq. (\ref{bve}) admits a solution of the form
\begin{eqnarray}
f({\mathbf r},{\mathbf k},t)&=&f_0({\mathbf R}\left(t\right),{\mathbf K}\left(t\right)),\label{scal}
\end{eqnarray}
where $f_0({\mathbf r},{\mathbf k})\equiv f({\mathbf r},{\mathbf k},t=0)$ represents the equilibrium PSDF and
\begin{eqnarray}
R_\eta(t)&=&\frac{r_\eta}{b_\eta(t)},\label{scal1}\\
K_\eta(t)&=&b_\eta(t)k_\eta-\frac{M}{\hbar}\dot b_\eta(t)r_\eta,\label{scal2}
\end{eqnarray}
with $b_\eta$ being the time-dependent scaling parameters. Previously, this scaling ansatz has been widely adopted to study the dynamical properties of the quantum gases~\cite{he,sogo,castin,kagan,odelin,stringari1,liu1,liu2}. Alternatively, Lima and Pelster~\cite{lima,lima2} have presented a variational time-dependent Hartree-Fock theory for dipolar Fermi gas by employing a common-phase approximation. We show that the scaling ansatz, Eqs.~(\ref{scal})-(\ref{scal2}), can be derived by adopting a harmonic ansatz for the common phase in App.~\ref{appa}.

After introducing the scaling ansatz, the time dependence of the system is completely characterized by three scaling parameters $b_\eta(t)$. Following the standard procedure, it can be shown that the scaling parameters obey the following coupled dynamical equations
\begin{eqnarray}
\ddot b_\eta+\omega_\eta^2b_\eta-\frac{\hbar^2\left\langle K_\eta^2\right\rangle_0}{M^2b_\eta^3\left\langle R_\eta^2\right\rangle_0}+\frac{{\mathcal T}_{\eta}^{H}({\mathbf b})+{\mathcal T}_{\eta}^{F}({\mathbf b})}{Mb_\eta\left\langle R_\eta^2\right\rangle_0}=0,\label{b-eq}
\end{eqnarray}
where $\left\langle K_\eta^2\right\rangle_0=\left(2\pi\right)^{-3}\int d{\mathbf R}d{\mathbf K}K_\eta^2f_0\left({\mathbf R},{\mathbf K}\right)$ and $\left\langle R_\eta^2\right\rangle_0=\left(2\pi\right)^{-3}\int d{\mathbf R}d{\mathbf K}R_\eta^2f_0\left({\mathbf R},{\mathbf K}\right)$ represent, respectively, the average sizes of the equilibrium cloud in momentum and real spaces. The second and third terms on the left-hand-side of Eq.~(\ref{b-eq}) originate from the external trapping potential and kinetic terms, respectively. The contribution from Hartree-direct interaction is
\begin{eqnarray}
{\mathcal T}_\eta^H\left({\mathbf b}\right)=\frac{1}{2}\int\frac{d{\mathbf P}}{\left(2\pi\right)^3}\widetilde W_\eta\left({\mathbf b};{\mathbf P}\right)\tilde n_0\left({\mathbf P}\right)\tilde n_0\left(-{\mathbf P}\right),
\end{eqnarray}
where $n_0({\mathbf R})=(2\pi)^{-3}\int d{\mathbf K}f_0({\mathbf R},{\mathbf K})$ and
\begin{eqnarray}
\tilde n_0({\mathbf P})&=&{\mathcal F}\left[n_0({\mathbf R})\right],\nonumber\\
\widetilde W_\eta\left({\mathbf b};{\mathbf K}\right)&=&{\mathcal F}\left[R_\eta\frac{\partial W\left({\mathbf b};{\mathbf R}\right)}{\partial R_\eta}\right],\nonumber
\end{eqnarray}
with ${\mathcal F}\left[\cdot\right]$ denoting the Fourier transform and 
\begin{eqnarray}
W\left({\mathbf b};{\mathbf R}\right)\equiv c_d\frac{b_x^2X^2+b_y^2Y^2-2b_z^2Z^2}{\left(b_x^2X^2+b_y^2Y^2+b_z^2Z^2\right)^{5/2}}
\end{eqnarray}
being the dipolar interaction potential under the scaling transformation. Finally, the term corresponding to Fock-exchange interaction is
\begin{eqnarray}
{\mathcal T}_\eta^F\left({\mathbf b}\right)&=&-\frac{1}{2}\int\frac{d{\mathbf R}d{\mathbf K}d{\mathbf K}'}{\left(2\pi\right)^6}\widetilde W_\eta\left({\mathbf b};{\mathbf K}-{\mathbf K}'\right)\nonumber\\
&&\times f_0\left({\mathbf R},{\mathbf K}'\right)f_0\left({\mathbf R},{\mathbf K}\right).
\end{eqnarray}
Once the equilibrium distribution $f_0({\mathbf R},{\mathbf K})$ is known, the values of ${\cal T}_\eta^H({\mathbf b})$ and ${\cal T}_\eta^F({\mathbf b})$ can be calculated through numerical integrations for any given ${\mathbf b}$.

One can also study the collective excitations of a trapped dipolar Fermi gas by linearizing the dynamical equations~(\ref{b-eq}) around an equilibrium distribution (${\mathbf b}={\mathbf 1}$). To this end, we assume that 
\begin{eqnarray}
b_\eta=1+\tilde b_\eta\label{linear}
\end{eqnarray}
with $\tilde b_\eta$ being small quantities. Substituting Eqs.~(\ref{linear}) into Eqs.~({\ref{b-eq}) and keeping the linear terms in $\tilde b_j$, we find
\begin{eqnarray}
\ddot{\tilde b}_\eta+\left(\omega_\eta^2+\frac{3\hbar^2\langle K_\eta^2\rangle_0}{M^2\langle R_\eta^2\rangle_0}\right)\tilde b_\eta+\sum_{\eta'}\frac{{\mathcal S}_{\eta\eta'}\tilde b_{\eta'}}{2M\langle R_{\eta'}^2\rangle_0}=0,\label{lineq}
\end{eqnarray}
where
\begin{eqnarray*}
{\mathcal S}_{\eta\eta'}&=&\int d{\mathbf r}d{\mathbf r}'V_d^{(\eta\eta')}\left({\mathbf r}-{\mathbf r}'\right)n_0\left({\mathbf r}'\right)n_0\left({\mathbf r}\right),\\
&&-\int\frac{d{\mathbf r}d{\mathbf k}d{\mathbf k}'}{\left(2\pi\right)^6}\widetilde V_d^{(\eta\eta')}\left({\mathbf k}-{\mathbf k}'\right)f_0\left({\mathbf r},{\mathbf k}'\right)f_0\left({\mathbf r},{\mathbf k}\right)
\end{eqnarray*}
with $V_d^{(\eta\eta')}\left({\mathbf r}\right)=
r_\eta r_{\eta'}\partial^2V_d\left({\mathbf r}\right)/(\partial r_\eta\partial r_{\eta'})$
and  $\widetilde V_d^{(\eta\eta')}\left({\mathbf k}\right)={\mathcal F}\left[V_d^{(\eta\eta')}\left({\mathbf r}\right)\right]$. 

We then look for the stationary solutions of the form 
\begin{eqnarray}
\tilde b_\eta\left(t\right)=\tilde b_\eta\left(0\right)\exp\left(-i\Omega t\right). \label{stso}
\end{eqnarray}
Submitting Eq.~(\ref{stso}) into the differential equations (\ref{lineq}), we obtain a system of linear equations whose non-trivial solutions are determined by the characteristic equation
\begin{eqnarray}
\left\vert\begin{array}{ccc} -\Omega^2+a_{xx} & a_{xy} & a_{xz} \\
a_{yx} & -\Omega^2+a_{yy} & a_{yz} \\ a_{zx} & a_{zy} & -\Omega^2+a_{zz} \end{array} \right\vert =0,
\end{eqnarray}
where the matrix elements are defined as
\begin{eqnarray}
a_{\eta\eta'}=\left[4\omega_\eta^2+\frac{3{\cal T}_\eta}{2M\left\langle R^2_\eta\right\rangle_0}\right]\delta_{\eta\eta'}+\frac{{\mathcal S}_{\eta\eta'}}{2M\left\langle R^2_\eta\right\rangle_0}\label{aee}
\end{eqnarray}
with ${\cal T}_\eta={\cal T}^{(H)}_\eta({\mathbf 1})+{\cal T}^{(F)}_\eta({\mathbf 1})$. Utilizing the axial symmetry of the system, one finds $a_{xx}=a_{yy}$, $a_{xy}=a_{yx}$, $a_{xz}=a_{yz}$, and $a_{zx}=a_{zy}$, which allows us to reduce the number of independent matrix elements.

\begin{figure}[tbp]
\centering
\includegraphics[width=3.in]{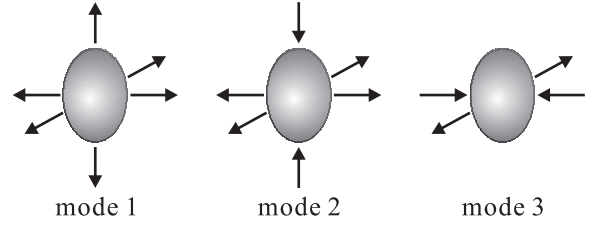}
\caption{(Color online) Schematical plot of the three shape oscillation modes.}
\label{modes}
\end{figure}

Solving the characteristic equation analytically, we find three eigenfrequencies
\begin{eqnarray}
\Omega_{1,2}^2&=&\frac{1}{2}\left(a_{xx}+a_{xy}+a_{zz}\pm\sqrt{\mathcal A}\right),\label{eq:freq1}\\
\Omega_3^2&=&a_{xx}-a_{yx}\label{eq:freq}
\end{eqnarray}
where ${\mathcal A}=\left(a_{xx}+a_{xy}-a_{zz}\right)^2+8a_{xz}a_{zx}$. The corresponding unnormalized eigenvectors are
\begin{eqnarray*}
{\mathbf u}_{1,2}=\left(v_{1,2},v_{1,2},1\right)^T \mbox{ and } {\mathbf u}_3=\left(-1,1,0\right)^T\label{eigenv}
\end{eqnarray*}
with $v_{1,2}=\left(a_{xx}+a_{xy}-a_{zz}\pm\sqrt{\mathcal A}\right)/(4a_{zx})$. In Fig.~\ref{modes}, we graphically illustrate three shape oscillation modes. In a spherically symmetric system, a collective oscillation mode can be uniquely characterized by the principal ($n$), azimuthal ($l$), and magnetic ($m$) quantum numbers. Since our system only possesses an axial symmetry, $n$ and $l$ are no longer good quantum numbers. Therefore, modes 1 and 2 result from the coupling of the monopole ($n=1,l=0,m=0$) and quadrupole ($n=0,l=2,m=0$) modes; While mode 3 represents the $m=2$ quadrupole mode. 

In App.~\ref{apposc}, we present an alternative calculation for the frequencies of the three shape oscillation modes based on the sum-rule approach. We analytically show that these two approaches generate the same oscillation frequencies.

\section{results}\label{resu}
To present our results, we introduce a set of dimensionless units based on the geometric average of the trap frequencies $\bar\omega=(\omega_\perp^2\omega_z)^{1/3}$ and the harmonic oscillator length $\bar a=\sqrt{\hbar/(M\bar\omega)}$: $N^{1/6}\bar a$ for length, $N^{1/6}\bar a^{-1}$ for wave vector, and $N^{1/3}\hbar\bar\omega$ for energy. Under these choices, the dipolar interaction strength is characterized by a dimensionless quantity
\begin{eqnarray}
D =\frac{N^{1/6}c_d}{\hbar\bar\omega \bar a^3}.
\end{eqnarray}
In a recent experiment~\cite{dwang}, a gas of $N\simeq 3.9\times 10^4$ ground state KRb molecules is realized. The transverse and axial frequencies of trapping potential are $\omega_\perp=(2\pi)\,32$ Hz and $\omega_z=(2\pi)\,195$ Hz, respectively, resulting in a pancake-shaped trap potential with aspect ratio $\lambda\simeq 6.1$. From these experimental parameters, one finds the dimensionless dipolar interaction strength to be
\begin{eqnarray}
D\simeq 9.62\bar d^2,
\end{eqnarray}
where $\bar d$ is the electric dipole moment in units of Debye. For rotating polar molecules, the value of $\bar d$ is tunable by an external electric field, however it is limited by the permanent dipole moment ($0.57$ Debye for the ground state KRb molecule). Therefore, the maximal value of $D$ can be realized in this system is around $3$. Finally, the temperature of the gas realized in the experiment is $T\simeq 220\,{\rm nK}\simeq1.3T_F$, where $T_F\equiv(6N)^{1/3}\hbar\bar\omega/k_B$ is the Fermi temperature.

Now, our system is completely specified by three parameters: the trap aspect ratio $\lambda$, the dimensionless interaction parameter $D$, and the temperature $T$~\cite{zhang2}. In the following, we investigate the dynamical properties of a dipolar Fermi gas, including free expansion, collapse dynamics, and low-lying collection excitations. To this end, we first obtain an equilibrium PSDF $f_0({\mathbf R},{\mathbf K})$ via the iterative procedure described in Sec.~\ref{equil}. We then numerically evolve Eqs.~(\ref{b-eq}) with the initial conditions $b_\eta(0)=1$ and $\dot b_\eta(0)=0$. Based on the scaling ansatz, the PSDF $f({\mathbf r},{\mathbf k},t)$ can then be found from Eqs.~(\ref{scal})-(\ref{scal2}), which allows us to calculate various physical quantities.

\subsection{Free expansion}\label{frexp}
\begin{figure}[tbp]
\centering
\includegraphics[width=3.3in]{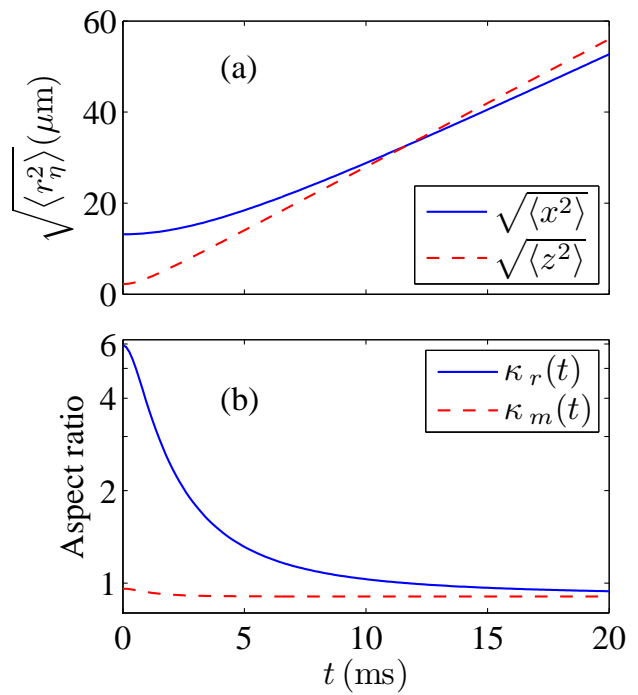}
\caption{(Color online) Time dependences of the rms cloud widths (a) and the aspect ratios of the distributions in real and momentum spaces (b) for $N=3.9\times 10^4$, $\lambda=6.1$, $\bar\omega=(2\pi)\,58.4\,$Hz, $D=2$, and $T=0.55T_F$.}
\label{trmsab}
\end{figure}

Let us first consider the free expansion of an initially trapped dipolar Fermi gas. As a diagnostic tool, the time-of-flight imaging has been used extensively in cold atom physics. The expanded cloud directly reflects the momentum distribution of the system, which bears the signature of the underlying dipolar interaction. In fact, the dipolar effect in Cr condensates was first detected by comparing the aspect ratios of the expand clouds~\cite{pfau}. Particularly, for a gas of KRb molecules, D. Wang {\it et al.}~\cite{dwang} have experimentally realized a scheme for direct absorption imaging of an ultracold polar molecular gas at arbitrary external electric or magnetic field. 

The expansion dynamics can be studied by removing the restoring force terms $\omega_\eta^2 b_\eta$ in Eqs.~(\ref{b-eq}), which corresponds to turning off the external trapping potential. In terms of the scaling parameters $b_\eta$, the root-mean-square (rms) cloud widths are
\begin{eqnarray}
\sqrt{\langle r_\eta^2\rangle}=b_\eta(t)\sqrt{\langle R_\eta^2\rangle_0}.
\end{eqnarray}
In particular, if one switches off the inter-particle dipolar interaction and lets the cloud expand ballistically, equations~(\ref{b-eq}) can be solved analytically to yield
\begin{eqnarray}
b^{(0)}_\eta\left(t\right)=\sqrt{1+M^{-2}\hbar^2t^2\left\langle K_\eta^2\right\rangle_0/\left\langle R_\eta^2\right\rangle_0}.\label{ballb}
\end{eqnarray}
We point out that, for rotating polar molecules, one can easily switches off the dipolar interaction by removing the external electric field. In Fig.~\ref{trmsab}(a), we present the typical results for the rms cloud widths as functions of time for a dipolar Fermi gas initially trapped in a pancake-shaped potential using experimental parameters. It can be seen that, similar to the zero-temperature case~\cite{he}, the real-space density distribution eventually becomes stretched along $z$-axis during the expansion due to the anisotropic dipolar interaction. As we shall show, this conclusion is independent of the trap geometry.

\begin{figure}[tbp]
\centering
\includegraphics[width=3.3in]{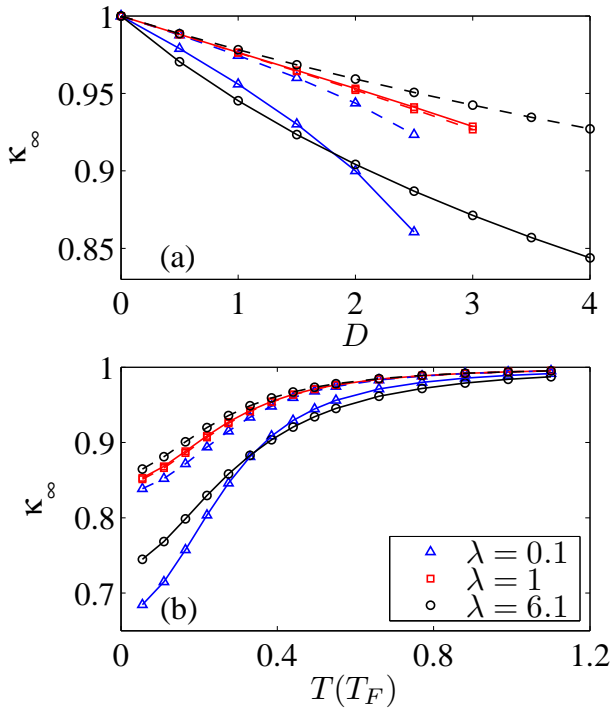}
\caption{(Color online) Dipolar interaction strength (a) and temperature (b) dependences of the asymptotic aspect ratio $\kappa_\infty$ of the cloud for various trap geometries. The solid and dashed lines correspond to, respectively, the free and ballistic expansions. Other parameters are $T/T_F=0.55$ in (a) and $D=1$ in (b).}
\label{binfty}
\end{figure}

The axial symmetry of the system implies $b_x=b_y\equiv b_\perp$, which allows us to characterize the time-dependent deformations of the distributions using
\begin{eqnarray}
\kappa_r(t)=\sqrt{\langle x^2\rangle/\langle z^2\rangle}\;\mbox{ and }\; \kappa_m(t)=\sqrt{\langle k_x^2\rangle/\langle k_z^2\rangle}\nonumber
\end{eqnarray}
in the real and momentum spaces, respectively. For an initial equilibrium distribution, $\kappa_r(0)$ strongly depends on the trap geometry (In fact, for a trapped ideal Fermi gas, it can be easily verified that, independent of the temperature, the initial real space aspect ratio is exactly the trap aspect ratio $\lambda$); while, as a result of the Fock-exchange interaction, the initial momentum distribution is always stretched along $z$-axis [$\kappa_m(0)<1$]. For the dynamical behaviors, as shown in Fig.~\ref{trmsab}(b), both $\kappa_r(t)$ and $\kappa_m(t)$ approach the same asymptotic value $\kappa_\infty$ at large $t$ limit. This can be easily understood by noting that the shape of the expanding cloud is essentially determined by the momentum distribution when the dipolar interaction is negligible. In particular, for ballistic expansion during which the inter-particle interaction is absent, it can be easily verified from Eq. (\ref{ballb}) that $\kappa_\infty=\kappa_m(0)$.

In Fig.~\ref{binfty}, we plot the dipolar interaction strength and temperature dependences of the asymptotic aspect ratio $\kappa_\infty$ for various initial trap geometries. As a comparison, the initial momentum-space deformation $\kappa_m(0)$ is also plotted. A general observation is that $\kappa_\infty<1$ under all situations, indicating that the expanding cloud eventually becomes cigar-shaped. In addition, increasing interaction strength or lowering temperature results in larger anisotropy of the expanding cloud. During free expansion, both Hartree-direct and Fock-exchange interaction energies are eventually converted into the kinetic energy. As the direct dipolar interaction tends to stretch the expanding cloud along $z$-axis, one finds that $\kappa_\infty$ is significantly smaller than $\kappa_k(0)$ in a highly anisotropic trap. However, in a spherical trap, where the direct dipolar interaction is small, the discrepancy between $\kappa_\infty$ and $\kappa_m(0)$ is negligible.

\subsection{Collapse dynamics}\label{coldyn}
Now, we turn to study the collapse dynamics of a trapped dipolar Fermi gas. A Fermi gas becomes unstable when the degenerate pressure is unable to balance the attractive inter-particle interaction. The collapses of an ultracold Fermi gas was demonstrated in boson-fermion mixtures~\cite{Collapse_Bose_Fermi_mixture}, in which the instability of the Fermi gas is induced by the attractive interaction between the bosons and fermions. However, for a two-component Fermi gas with interspecies $s$-wave interaction, collapse has not yet been achieved experimentally. In fact, as the total energy remains positive~\cite{gehm,heis}, the system is stable even in the unitary regime where the scattering length is negative infinity. Only at very high density, when the inter-particle spacing becomes comparable to the effective range of the interaction, the system can in principle becomes unstable~\cite{fregoso}. 

On the other hand, with the long-range and partially attractive inter-particle interaction, the dipolar Fermi gases may provide a prospective platform to study the collapse dynamics of the fermionic gases. Calculations within mean-field have shown that, for given $\lambda$ and $T$, there exists a critical dipolar interaction strength $D^*$ such that the system becomes unstable when $D>D^*$~\cite{zhang2}.

To reveal the dynamical process of a collapse, we first prepare an initially stable state under a given set of control parameters: $\lambda$, $T$, and $D=D_i$. At time $t=0^+$, the dipolar interaction strength $D$ is suddenly increased to $D_f$ ($>D^*$). We then numerically evolve Eqs.~(\ref{b-eq}) to simulate the dynamics of the system. In terms of $b_\eta$, collapses occur when at least one of the scaling parameters approaches zero. Additionally, we say that a collapse is isotropic if all scaling parameters go to zero simultaneously; otherwise, it is anisotropic. 

\begin{figure}[tbp]
\centering
\includegraphics[width=2.8in]{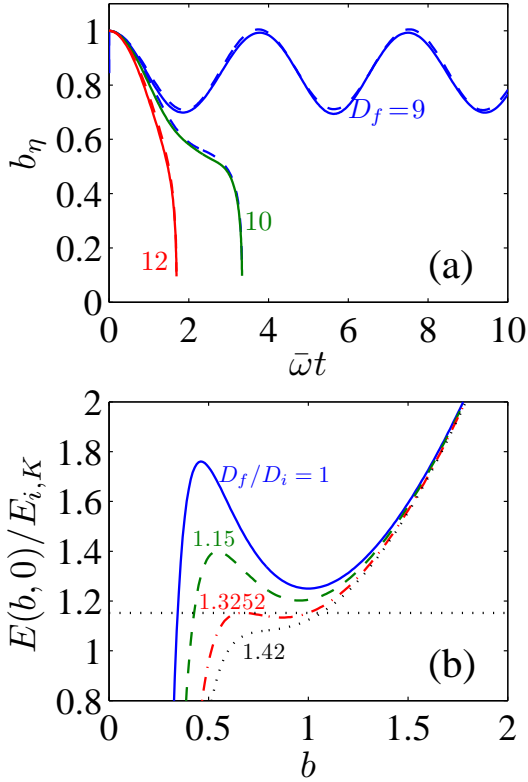}
\caption{(Color online) (a) Typical dynamical behaviors of $b_\perp$ (solid lines) and $b_z$ (dashed lines) for $\lambda=1$, $T=0.28T_F$, $D_i=2$, and various $D_f$'s. (b) $E(b,0)/E_{i,K}$ as a function of $b$ for $\gamma_i=-0.3$ and various $D_f/D_i$'s. The horizontal dotted line denotes the total energy $E(1,0)/E_{i,K}$ for $D_f/D_i=1.3252$.}\label{colsph}
\end{figure}

Figure~\ref{colsph}(a) plots the typical dynamical behaviors of the scaling parameters for $\lambda=1$, $T=0.28T_F$, and after the dipolar interaction strength is suddenly increased from $D_i=2$ to various $D_f$'s. As can be seen, the system becomes unstable and starts to collapse isotropically only when $D_f$ is larger than a threshold value $D_f^*\simeq 10$. Surprisingly, we find that $D_f^*$ is significantly larger than $D^*=2.04$ for the equilibrium state. To find an explanation, we assume that $b_\perp(t)\simeq b_z(t)\equiv b(t)$ as shown in Fig.~\ref{colsph}(a). As a result, the total energy for $t\geq 0^+$ becomes
\begin{eqnarray}
E(b,\dot b)=\frac{E_{i,K}}{b^2}+E_{i,P}b^2+\frac{D_f}{D_i}\frac{E_{i,I}}{b^3}+\dot b^2\sum_\eta\frac{1}{2}M\langle R_\eta^2\rangle_0,\nonumber\\\label{etot}
\end{eqnarray}
where $E_{i,K}=(2\pi)^{-3}\int d{\mathbf r}d{\mathbf k}\left(\hbar^2{\mathbf k}^2/2M\right)f_0({\mathbf r},{\mathbf k})$, $E_{i,P}=(2\pi)^{-3}\int d{\mathbf r}d{\mathbf k}U_{\rm ho}({\mathbf r})f_0({\mathbf r},{\mathbf k})$, and
\begin{eqnarray}
E_{i,I}&=&\frac{1}{2}\int d{\mathbf r}d{\mathbf r}'n_0({\mathbf r})n_0({\mathbf r}')V_d({\mathbf r}-{\mathbf r}')\nonumber\\
&&-\frac{1}{2}\int \frac{d{\mathbf r}d{\mathbf k}d{\mathbf k}'}{(2\pi)^6}f_0({\mathbf r},{\mathbf k})f_0({\mathbf r},{\mathbf k}')\widetilde V_d({\mathbf k}-{\mathbf k}')\nonumber
\end{eqnarray}
are, respectively, the kinetic, potential, and interaction energies of the equilibrium state (with dipolar interaction strength $D_i$). 

Let us first consider the initial state by taking $D_f=D_i$ in Eq.~(\ref{etot}). Consequently, the total energy is denoted as $E_i$. The stationary condition, $\frac{\partial E_i}{\partial b}|_{b=1,\dot b=0}=0$, yields 
\begin{eqnarray}
2E_{i,K}-2E_{i,P}+3E_{i,I}&=&0,\nonumber
\end{eqnarray}
which is exactly the Virial theorem. Furthermore, the stability condition for the initial state, $\frac{\partial^2 E_i}{\partial b^2}|_{b=1,\dot b=0}\geq 0$, requires
\begin{eqnarray}
\gamma_i\equiv\frac{E_{i,I}}{E_{i,K}}\geq \gamma^*\equiv-\frac{8}{15},\nonumber
\end{eqnarray}
indicating that the ratio of the interaction energy to the kinetic energy must be larger than a critical value $\gamma^*$ for the initial state.

After the dipolar interaction strength is switched to $D_f$, the stability of the system can then be analyzed by examining the $b$-dependence of $E(b,\dot b)$ at $t=0^+$, i.e.,
\begin{eqnarray}
\frac{E(b,0)}{E_{i,K}}=\frac{1}{b^2}+\left(1+\frac{3\gamma_i}{2}\right)b^2 +\frac{D_f}{D_i}\frac{\gamma_i}{b^3}.\label{tote}
\end{eqnarray}
Apparently, the system is always stable if $\gamma_i\geq 0$. Therefore, in order to induce a collapse, one must have $\gamma_i<0$ for the initial state. Moreover, to find the threshold $D_f^*$, we plot, in Fig.~\ref{colsph}(b), the typical behaviors of $E(b,0)$ corresponding to $\gamma_i=-0.3$ and various $D_f$'s. As can be seen, when $D_f/D_i$ is not very large, $E(b,0)$ has a local maximum at $b=b_{\rm max}$, with $b_{\rm max}$ being determined by the conditions $\left.\frac{\partial E(b,0)}{\partial b}\right|_{b=b_{\rm max}}=0$ and $\left.\frac{\partial^2 E(b,0)}{\partial b^2}\right|_{b=b_{\rm max}}\leq 0$. However, this local maximum vanishes for sufficiently large $D_f/D_i$, under which the system becomes unstable. From above analysis, it becomes clear that the equation 
\begin{eqnarray}
E(1,0)=E(b_{\rm max},0)\label{thresh}
\end{eqnarray}
should be satisfied when the threshold $D_f^*$ is reached. For the example in Fig.~\ref{colsph}(a), we have $\gamma_i=-0.0692$. A threshold $D_f^*=9.68$ can then be determined from Eq.~(\ref{thresh}), in agreement with our numerical finding.

To quantitatively show that $D_f^*$ is larger than $D^*$, we need to find the relation between $\gamma_i$ and $D_i$ numerically. Here, for simplicity, we present a qualitative argument. To this end, we consider two limit cases with $\gamma_i\rightarrow 0^-$ and $\gamma^*$, for which the threshold dipolar interaction strengths are, respectively, $D_f^*=\infty$ and $D^*$. Since $D_f^*\geq D^*$ in both cases, as a natural generalization, it should be held for any $\gamma\in(\gamma^*,0)$.

\begin{figure}[tbp]
\centering
\includegraphics[width=3.3in]{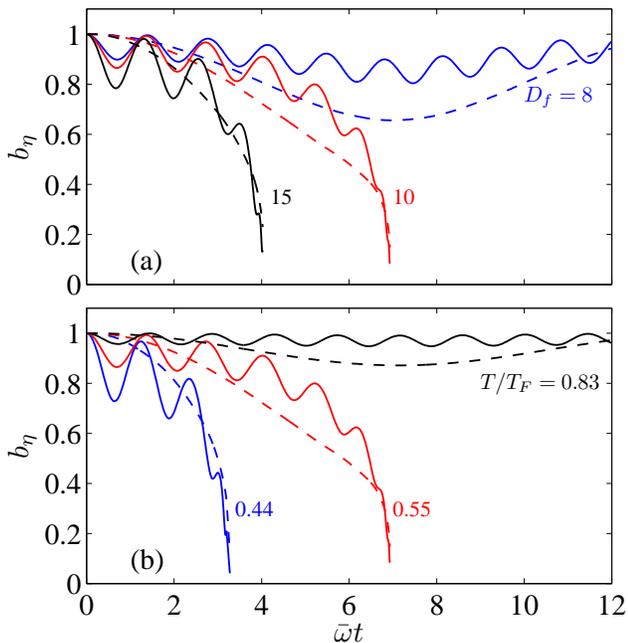}
\caption{(Color online) Typical dynamical behaviors of $b_\perp$ (solid lines) and $b_z$ (dashed lines) in a cigar-shaped trap ($\lambda=0.1$) after the dipolar interaction strength is suddenly increased from $D_i=2$ to $D_f$. (a) $D_f$ dependence for a given temperature $T=0.55T_F$. (b) Temperature dependence for given $D_f=10$.}\label{fig:collapse_time}
\end{figure}

Figure~\ref{fig:collapse_time}(a) shows the typical dynamical behaviors of $b_\eta$ for $\lambda=0.1$, $T=0.55T_F$, $D_i=2$, and $D_f$'s. As can be seen, there also exists a threshold $D_f^*$ in a cigar-shaped trap. In addition, the system collapses faster with a larger $D_f$. In Fig.~\ref{fig:collapse_time}(b), for the fixed $D_i=2$ and $D_f=10$, we compare the collapse dynamics by varying the temperature of the system. For the high temperature case ($T=0.83T_F$), the equilibrium PSDF of the trapped dipolar Fermi gas becomes more resembling of that of a trapped ideal Fermi gas, such that the dipolar interaction energy is negligible small under given $D_i$. Consequently, the system remains dynamically stable. However, collapses are realized when the temperature is lowered.

An important feature revealed in Fig.~\ref{fig:collapse_time} is that, whenever the system becomes dynamically unstable in a cigar-shaped trap, the scaling parameters $b_\perp$ and $b_z$ always go to zero simultaneously, suggesting that a dipolar Fermi gas always undergoes isotropic collapses. This feature is in striking difference with the anisotropic collapse of a dipolar condensate~\cite{ODell_col}, and it is caused by the Fock-exchange interaction. In fact, if we artificially remove ${\mathcal T}_j^F\left({\mathbf b}\right)$ term when evolving Eqs.~(\ref{b-eq}), the collapse becomes anisotropic again.

In highly pancake-shaped traps, the system remains dynamically stable for all control parameters we have tested. This can be easily understood based on our argument for the spherical trap case. Indeed, we find that the total interaction energy for an equilibrium state is always positive. Even though the variational calculation shows that the interaction energy may eventually becomes negative for sufficiently large $D_i$, the full numerical calculation, on the other hand, indicates that the system collapses locally under such dipolar interaction strength.

%
\subsection{Collective excitations}\label{colex}
Finally, we consider the low-lying collective excitations of a trapped dipolar Fermi gas, which also requires knowing the equilibrium distribution function for a given set of control parameters. In the non-interacting limit, these oscillation frequencies reduce to
$\Omega_{1,2}^{(0)}=\sqrt{2}\omega_\perp\sqrt{1+\lambda^2\pm|\lambda^2-1|}$ and $\Omega_3^{(0)}=2\omega_\perp$, which are independent of the temperature. After the interaction is switched on, the collective excitation frequencies will be shifted with respect to $\Omega_i^{(0)}$. For the same $D$ value, the interaction induced frequency shifts in highly anisotropic traps are usually larger than those in a spherical trap, due to the strong direct dipolar interaction. Therefore, we will concentrate on the low-lying collective excitations in cigar- and pancake-shaped traps.

\begin{figure}[tbp]
\centering
\includegraphics[width=3.in]{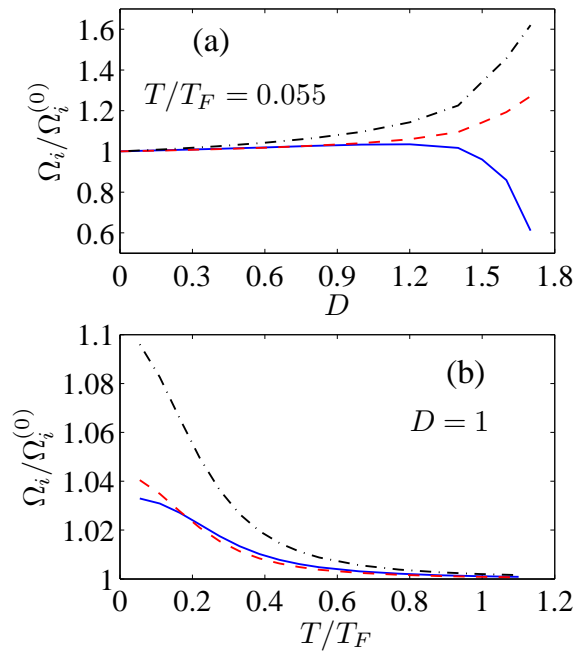}
\caption{(Color online) Dipolar interaction strength (a) and temperature (b) dependences of the shape oscillation frequencies for a cigar-shaped trap with $\lambda=0.1$. Modes 1, 2, and 3 are labeled with solid, dashed, and dash-dotted lines, respectively.}
\label{fig:ce_cigar}
\end{figure}

Figure~\ref{fig:ce_cigar}(a) shows the dipolar interaction dependence of the shape oscillation frequencies for $\lambda=0.1$ and $T=0.055T_F$. For relatively weak dipolar interaction, the frequencies of all three modes are slightly shifted upwards, whereas, the frequency of the mode 1 (monopole) starts to go down for $D>1.2$. Close to the stability boundary, the value of $\Omega_1$ drops significantly. The fact that the monopole mode goes soft in strong dipolar interaction regime is in agreement with  the variational calculation at zero-temperature limit~\cite{sogo}. Moreover, it is also consistent with the isotropic collapse in a cigar-shaped trap discussed previously. We remark that the softening of the monopole mode sensitively depends on temperature of the system. For instance, when the temperature is increased to $T=0.55T_F$, the frequencies of the three modes shift all the way upwards. This can be understood by examining the equilibrium distribution function at the vicinity of $D^*$, from which one may identify the type of the instability when the system becomes unstable. In fact, for the low temperature case ($T=0.055T_F$), the volume of the equilibrium gas goes to zero when $D$ approaches $D^*$, indicating that the system undergoes a global collapse. However, for the high temperature case ($T=0.55T_F$), local collapse is induced at the onset of the instability, which is not described by the three shape oscillation modes discussed here.

In Fig.~\ref{fig:ce_cigar}(b), we plot $\Omega_i$ as functions of $T$ for $\lambda=0.1$ and $D=1$. Among the three shape oscillation modes, the frequency of the mode 3 has the largest deviation from that of an ideal gas, for which the dipolar interaction induced frequency shift can be as high as $10\%$ at the low temperature limit. When the temperature is increased, the frequencies of all shape oscillation modes shift downward and asymptotically approach those of a non-interacting gas.

\begin{figure}[tbp]
\centering
\includegraphics[width=3.in]{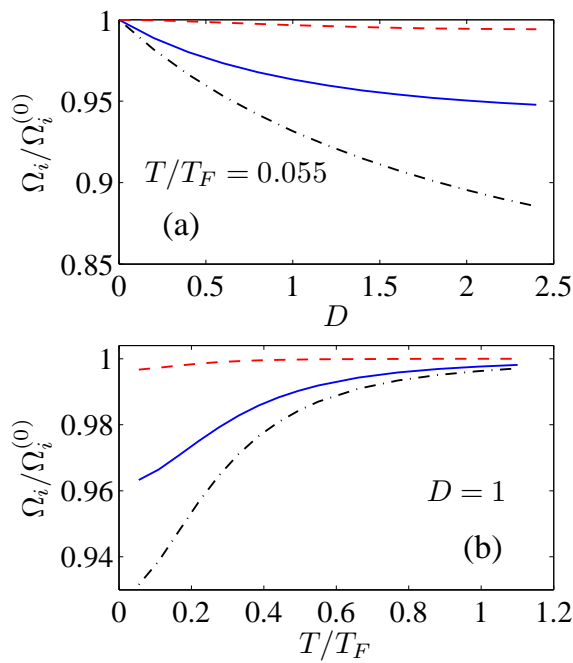}
\caption{(Color online) Same as Fig.~\ref{fig:ce_cigar} except for $\lambda=6.1$.}
\label{fig:ce_cake}
\end{figure}

For pancake-shaped traps ($\lambda=6.1$), Figure~\ref{fig:ce_cake}(a) shows the $D$ dependences of the shape oscillation frequencies. Taking into account the rescaling frequencies $\Omega_i^{(0)}$, the highest mode has a breathing geometry (mode 1), whereas the lowest one is the $m=2$ quadrupole mode. Even though the frequencies of all shape oscillation modes shift downwards, we do not find the frequency of any particular mode drops significantly close to the instability, which is the manifestation of the local collapses for the given trap geometry. The temperature dependences of the mode frequencies are presented in Fig.~\ref{fig:ce_cake}(b). As one increases the temperature, $\Omega_i$ increase monotonically to approach the frequencies corresponding to an ideal gas, showing an opposite tendency compared to the cigar-shaped trap. Again, the $m=2$ quadrupole mode has the largest frequency shift with respect to an ideal gas.

\section{conclusions}\label{concl}
In this paper, we have explored the dynamical properties of a trapped dipolar Fermi gas at finite temperature. For free expansion, we find the expanded cloud always becomes stretched along the direction of the dipole moment. We further explore the temperature and interaction strength dependences of the asymptotical aspect ratio of the expanded cloud. We have also studied the collapse dynamics by suddenly increasing the dipolar interaction strength. In contrast to the anisotropic collapse of a dipolar condensate, we find that dipolar Fermi gases always collapse isotropically. Finally, we have investigated the low-energy shape oscillations of a trapped dipolar Fermi gas. It is shown that, in a cigar-shaped trap, the monopole mode goes soft close to the instability, which is consistent with the isotropic collapse of the dipolar Fermi gas. In addition, among three shape oscillation modes, the $m=2$ quadrupole mode always has the largest interaction induced frequency shift with respect to that of an ideal gas.

\section*{ACKNOWLEDGMENTS}
This work was supported by the NSFC (Grant Nos. 11025421, 10935010, and 10974209) and the ``Bairen" program of the Chinese Academy of Sciences.

\appendix
\section{Scaling ansatz and common phase approximation}\label{appa}
It is well-known that the phase-space distribution function $f({\mathbf r},{\mathbf k})$ is related to the single-particle reduced density matrix as
\begin{eqnarray}
f\left({\mathbf r},{\mathbf k}\right)=\int\frac{d{\mathbf k}}{\left(2\pi\right)^3}e^{i{\mathbf k}\cdot{\mathbf s}}\rho_1\left({\mathbf r}+\frac{\mathbf s}{2},{\mathbf r}-\frac{\mathbf s}{2}\right),\label{eq:psdf_def}
\end{eqnarray}
where $\rho_1\left({\mathbf r},{\mathbf r}'\right)={\rm Tr}\left[\hat\rho\hat\psi^\dag\left({\mathbf r}'\right)\hat\psi\left({\mathbf r}\right)\right]$ with $\hat\rho$ being the density operator and $\hat\psi$ the annihilation operator for fermionic field. To obtain an explicit expression for $\rho_1$, we assume that $\{\phi_a\}$ a set of single-particle orbitals, obtained by self-consistently solving the Hartree-Fock equations. For an equilibrium state, the single-particle reduced density matrix then becomes
\begin{eqnarray}
\rho_1\left({\mathbf r},{\mathbf r}'\right)=\sum_a\frac{\phi_a\left({\mathbf r}\right)\phi_a^*\left({\mathbf r}'\right)}{e^{(\varepsilon_a-\mu)/k_BT}+1},
\end{eqnarray}
where $\varepsilon_a$ is the energy of the $a$-th single-particle orbital. 

We remark that, from the above equation, one may derive the Fermi-Dirac distribution for a trapped Fermi gas, Eq.~(\ref{eq:Fermi_Dirac}), by adopting the local density approximation. To this end, we assume that, in the vicinity of a given space point, the system is treated as a homogeneous system with a spatially dependent chemical potential $\mu\left({\mathbf r}\right)=\mu-U_{\rm ho}\left({\mathbf r}\right)$. Thus the single-particle orbitals are taken to be plane waves with the subscription $a$ being replaced by the wave vector ${\mathbf k}$. Consequently, the single-particle reduced density matrix becomes
\begin{eqnarray}
\rho_1\left({\mathbf r},{\mathbf r}'\right)=\int\frac{d{\mathbf k}}{\left(2\pi\right)^3}\frac{e^{i{\mathbf k}\cdot({\mathbf r}-{\mathbf r}')}}{\exp\left[\left(\varepsilon(\frac{{\mathbf r}+{\mathbf r}'}{2},{\mathbf k})-\mu\right)/k_BT\right]+1},\nonumber\\\label{eq:rho1}
\end{eqnarray}
where the quasi-particle dispersion relation $\varepsilon\left({\mathbf r},{\mathbf k}\right)$ takes the same form as Eq.~(\ref{quasipot}). The Fourier transform of Eq.~(\ref{eq:rho1}) is exactly the Fermi-Dirac distribution Eq.~(\ref{eq:Fermi_Dirac}).

To study the collective motion of a zero-temperature dipolar Fermi gas, Lima and Pelster employed a common phase approximation for the single-particle orbitals in Ref.~\cite{lima,lima2}, which assumes that
\begin{eqnarray}
\phi_a({\mathbf r},t)=e^{iM\chi({\mathbf r},t)/\hbar}\tilde\phi_a({\mathbf r},t),
\end{eqnarray}
where the common phase $\chi\left({\mathbf r},t\right)$ represents the collective motion of the system. The common phase is further assumed to adopt a harmonic ansatz~\cite{lima}
\begin{eqnarray}
\chi({\mathbf r},t)=\frac{1}{2}\sum_\eta \beta_\eta(t)r_\eta^2.
\end{eqnarray}
Clearly, $\chi({\mathbf r},t)$ gives rise to a common velocity field of the form
\begin{eqnarray}
v_\eta({\mathbf r},t)=\beta_\eta(t)r_\eta.\label{vfield}
\end{eqnarray}

%

Without loss of generality, we may set $\beta_\eta=\dot b_\eta/b_\eta$. The velocity field, Eq.~(\ref{vfield}), can then be interpreted as a global dilatation of each single-particle orbital induced by the scaling transform Eq.~(\ref{scal1}). We can now rewrite the wave function for the orbitals as
\begin{eqnarray}
\phi_a\left({\mathbf r},t\right)=(b_xb_yb_z)^{-1/2}e^{iM\chi\left({\mathbf r},t\right)/\hbar}\phi_a\left({\mathbf R},0\right),
\end{eqnarray}
where $\chi=\frac{1}{2}\sum_\eta \dot b_\eta r_\eta^2/b_\eta$ and the factor $(b_xb_yb_z)^{-1/2}$ is introduced to maintain the normalization. Consequently, the time-dependent single-particle reduced density matrix becomes
\begin{eqnarray}
\rho_1\left({\mathbf r},{\mathbf r}',t\right)=\frac{e^{iM\left[\chi\left({\mathbf r},t\right)-\chi\left({\mathbf r}',t\right)\right]/\hbar}}{b_xb_yb_z}\rho_1\left({\mathbf R},{\mathbf R}',0\right).
\end{eqnarray}
Using Eq. (\ref{eq:psdf_def}), one finds the time-dependent phase-space distribution function
\begin{eqnarray}
f\left({\mathbf r},{\mathbf k},t\right)
&=&\int d{\mathbf S}\exp\left[-i\sum_\eta\left(b_\eta k_\eta-M\hbar^{-1}\dot b_\eta r_\eta\right)S_\eta\right]\nonumber\\
&&\times\rho_1\left({\mathbf R}+\frac{\mathbf S}{2},{\mathbf R}-\frac{\mathbf S}{2},0\right)\nonumber\\
&=&f({\mathbf R},{\mathbf K},0),\label{comscal}
\end{eqnarray}
where ${\mathbf K}$ is defined by Eq. (\ref{scal2}). Apparently, Eq. (\ref{comscal}) is exactly the scaling ansatz introduced in Sec.~\ref{bvescal}.

\section{The sum-rule approach to collective excitations}\label{apposc}
Here we use the sum-rule approach to calculate the frequencies of the three shape oscillation modes~\cite{bohi,stringari2}. We assume that $E_n$ and $|E_n\rangle$ are, respectively, the eigenenergies and eigenstates of a many-body Hamiltonian $\hat H$. Moreover, $\hat F$ is an excitation operator of the system. In the sum-rule approach, the average excitation energies of the low-lying excitations, which correspond to different types of shape oscillations, can be obtained through
\begin{eqnarray*}
\hbar\Omega_{\hat F}=\sqrt{\frac{m_3}{m_1}},
\end{eqnarray*}
where
\begin{eqnarray*}
m_k=\sum_{n'\neq n}\left|\left\langle E_{n'}\left|\hat F\right|E_n\right\rangle\right|^2P_n\left(E_{n'}-E_n\right)^k
\end{eqnarray*}
is the $k$-th order moment of the strength function for the transition operator $\hat F$, with $P_n$ being the probability distribution of the eigenstate at given temperature $T$. Alternatively, we may express $m_1$ and $m_3$ as
\begin{eqnarray*}
m_1&=&\frac{1}{2}{\rm Tr}\left\{\hat\rho\left[\hat F^\dag,\left[\hat H,\hat F\right]\right]\right\},\\
m_3&=&\frac{1}{2}{\rm Tr}\left\{\hat\rho\left[\left[\hat F^\dag,\hat H\right],\left[\hat H,\left[\hat H,\hat F\right]\right]\right]\right\},
\end{eqnarray*}
where $\hat\rho=\sum_nP_n|E_n\rangle\langle E_n|$ is the density operator of the system.

In order to obtain the frequencies of the shape oscillations, we consider excitation operators
\begin{eqnarray*}
\hat F\left(n,l,m\right)=\sum_{i=1}^Nr_i^{2n+l}Y_{lm}\left(\theta_i,\phi_i\right),
\end{eqnarray*}
where $(r_i,\theta_i,\phi_i)$ is the spherical coordinate of the $i$-th particle. Since modes 1 and 2 result from the coupling of the monopole ($n=1,l=0,m=0$) and quadrupole ($n=0,l=2,m=0$) modes, we construct an excitation operator which is a linear combination of $\hat F(1,0,0)$ and $\hat F(0,2,0)$, i.e.,
\begin{eqnarray*}
\hat F_\alpha=\sum_{i=1}^N\left[\sin\alpha\left(x_i^2+y_i^2\right)+\cos\alpha z_i^2\right],
\end{eqnarray*}
where $\alpha\in\left[0,\pi\right)$ is a parameter to be determined. After a straightforward calculation, we obtain the average excitation frequency as a function of $\alpha$,
\begin{eqnarray}
\Omega_{\hat F_\alpha}(\alpha)=\sqrt{\frac{\xi_1\cos^2\alpha+\xi_2\sin\alpha\cos\alpha+\xi_3\sin^2\alpha} {\cos^2\alpha+\zeta \sin^2\alpha}},\label{omgf}
\end{eqnarray}
where $\xi_1=4\bar\omega^2\lambda^{4/3}+(3{\mathcal T}_z+{\mathcal S}_{zz})/(2M\left\langle z^2\right\rangle_0)$, $\xi_2=2{\mathcal S}_{xz}/(M\left\langle z^2\right\rangle_0)$, $\xi_3=8\bar\omega^2\lambda^{-2/3}\kappa_r^2\left(0\right) +(3{\mathcal T}_x+{\mathcal S}_{xx}+{\mathcal S}_{xy})/(M\left\langle z^2\right\rangle_0)$, and $\zeta=2\kappa_r^2\left(0\right)$.

To determine $\alpha$, we consider two general excitation operators, $\hat F_1$ and $\hat F_2$, with average excitation frequencies $\Omega_{\hat F_1}<\Omega_{\hat F_2}$. In addition, we assume that the modes excited by $\hat F_1$ and $\hat F_2$ are orthogonal. It can then be seen that, for an operator $\hat F$ constructed by superimposing $\hat F_1$ and $\hat F_2$, we always have $\Omega_{\hat F_1}\leq\Omega_{\hat F}\leq\Omega_{\hat F_2}$. Therefore, for {\em eigen-excitation} modes, such as those described by Eq. (\ref{eigenv}), the values of $\alpha$ should either maximize or minimize $\Omega_{\hat F_\alpha}(\alpha)$, which yields
\begin{eqnarray}
\cos\alpha_{1,2}=\frac{{\rm sign}(\xi_2)\left(\zeta \xi_1-\xi_3\pm\sqrt{\zeta \xi_2^2+\left(\zeta \xi_1-\xi_3\right)^2}\right)} {\sqrt{\xi_2^2+\left(\zeta\xi_1-\xi_3\pm\sqrt{\zeta \xi_2^2+\left(\zeta \xi_1-\xi_3\right)^2}\right)^2}}.\nonumber\\\label{alf}
\end{eqnarray}
Submitting Eq. (\ref{alf}) into (\ref{omgf}), we obtain the frequencies of the shape oscillation modes 1 and 2, which take exactly the same form as those in Eq.~(\ref{eq:freq1}).

The shape oscillation mode 3 can be directly excited by the transition operators $\hat F(0,2,\pm2)$. The average excitation frequencies are exactly Eq. (\ref{eq:freq}) and they are degenerate for $m=\pm2$ modes due to the axial symmetry of the system.


\begin{thebibliography}{}

\bibitem{ye1} K.-K. Ni, S. Ospelkaus, M. H. G. de Miranda, A. Pe'er, B. Neyenhuis, J.J. Zirbel, S. Kotochigova, P.S. Julienne, D.S Jin, and J. Ye, Science {\bf322}, 231 (2008).

\bibitem{ye2} S. Ospelkaus, K.-K. Ni, M. H. G. de Miranda, B. Neyenhuis, D. Wang, S. Kotochigova, P. S. Julienne, D. S. Jin, J. Ye, Faraday Discuss. {\bf142}, 351 (2009).

\bibitem{dwang} D. Wang, B. Neyenhuis, M. H. G. de Miranda, K.-K. Ni,
S. Ospelkaus, D. S. Jin, and J. Ye, Phys. Rev. A {\bf81}, 061404 (2010).

\bibitem{you} L. You and M. Marinescu, Phys. Rev. A {\bf 60}, 2324 (1999).

\bibitem{bara} M. A. Baranov, M. S. Mar'enko, Val. S. Rychkov, and G. V. Shlyapnikov, Phys. Rev. A {\bf66}, 013606 (2002).

\bibitem{bara2} M.A. Baranov, L. Dobrek, and M. Lewenstein, Phys. Rev. Lett. {\bf92}, 250403 (2004).

\bibitem{pu} C. Zhao L. Jiang, X. Liu, W. M. Liu, X. Zou, and H. Pu, Phys. Rev. A {\bf81}, 063642 (2010).

\bibitem{samo} K. V. Samokhin and M. S. Mar'enko, Phys. Rev. Lett. \textbf{97}, 197003 (2006).

\bibitem{wu} C. J. Wu and J. E. Hirsch, Phys. Rev. B \textbf{81}, 020508(R)
(2010).

\bibitem{shi} T. Shi, J.-N. Zhang, C.-P. Sun, and S. Yi, Phys. Rev. A {\bf82}, 033623 (2010).

\bibitem{bara3} M. A. Baranov, K. Osterloh, and M. Lewenstein, Phys. Rev. Lett. {\bf94}, 070404 (2005).

\bibitem{oste} K. Osterloh, N. Barber\'{a}n, and M. Lewenstein, Phys. Rev. Lett. {\bf99}, 160403 (2007).

\bibitem{liu} C. Lin, E. Zhao, and W.V. Liu, Phys. Rev. B {\bf81}, 045115 (2010).

\bibitem {quin} J. Quintanilla, S.T. Carr, and J.J. Betouras, Phys. Rev. A {\bf79}, 031601 (2009).

\bibitem{he2} L. He and W. Hofstetter, arXiv:1101.5633 (2011).

\bibitem{goral2} K. G\'{o}ral, B.-G. Englert, and K. Rz\c{a}\.{z}ewski, Phys. Rev. A {\bf 63}, 033606 (2001).

\bibitem{goral3} K. G\'{o}ral, M. Brewczyk, and K. Rz\c{a}\.{z}ewski, Phys. Rev. A {\bf67}, 025601 (2003).

\bibitem{he} L. He, J.-N. Zhang, Y.-B. Zhang, and S. Yi, Phys. Rev. A {\bf77} 031605, (2008).

\bibitem{miya} T. Miyakawa, T. Sogo, H. Pu, Phys. Rev. A {\bf77}, 061603 (2008).

\bibitem{zhang1} J.-N. Zhang and S. Yi, Phys. Rev. A {\bf80}, 053614 (2009).

\bibitem{chan} C.-K. Chan, C. Wu, W.-C. Lee, and S. Das Sarma, Phys. Rev. A {\bf81}, 023602 (2010). 

\bibitem{freg} B. M. Fregoso and E. Fradkin, Phys. Rev. Lett. {\bf 103}, 205301 (2009).

\bibitem{freg2} B. M. Fregoso, K. Sun, E. Fradkin, New J. Phys. {\bf 11}, 103003 (2009).

\bibitem{lin} C.-H. Lin, Y.-T. Hsu, H. Lee, and D.-W. Wang, Phys. Rev. A {\bf81}, 031601 (2010).

\bibitem{sogo} T. Sogo, L. He, T. Miyakawa, S. Yi, H. Lu, and H. Pu, New J.
Phys. {\bf11}, 055017 (2009).

\bibitem{lima} A.R.P. Lima and A. Pelster, Phys. Rev. A {\bf81}, 021606(R) (2010).

\bibitem{lima2} A.R.P. Lima, A. Pelster, Phys. Rev. A {\bf81}, 063629 (2010).

\bibitem{bohn} S. Ronen and J.L. Bohn, Phys. Rev. A {\bf81}, 033601 (2010).

\bibitem{zhang2} J.-N. Zhang and S. Yi, Phys. Rev. A {\bf81}, 033617 (2010).

\bibitem{bail} D. Baillie and P.B. Blakie, Phys. Rev. A {\bf 82}, 023605 (2010).

\bibitem{endo} Y. Endo, T. Miyakawa, and T. Nikuni, Phys. Rev. A {\bf 81}, 063624 (2010).

\bibitem{pfau} J. Stuhler, A. Griesmaier, T. Koch, M. Fattori, T. Pfau, S. Giovanazzi, P. Pedri, and L. Santos, Phys. Rev. Lett. {\bf95}, 150406 (2005).

\bibitem{bism} G. Bismut, B. Pasquiou, E. Mar¨¦chal, P. Pedri, L. Vernac, O. Gorceix, and B. Laburthe-Tolra, Phys. Rev. Lett. {\bf105}, 040404 (2010).

\bibitem{stringari1} C. Menotti, P. Pedri, and S. Stringari, Phys. Rev. Lett. {\bf89}, 250402 (2002).

\bibitem{castin} Y. Castin and R. Dum, Phys. Rev. Lett. {\bf77}, 5315 (1996). 

\bibitem{kagan} Y. Kagan, E. L. Surkov, and G. V. Shlyapnikov, Phys. Rev. A {\bf54}, R1753 (1996).

\bibitem{odelin} D. Gu\'{e}ry-Odelin, Phys. Rev. A {\bf66}, 033613 (2002).

\bibitem{liu1} X.-J. Liu and H. Hu, Phys. Rev. A {\bf67}, 023613 (2003). 

\bibitem{liu2} H. Hu, X.-J. Liu, and M. Modugno, Phys. Rev. A {\bf67}, 063614 (2003).

\bibitem{Collapse_Bose_Fermi_mixture} G. Modugno, G. Roati, F. Riboli, F. Ferlaino, R. J. Brecha, and M. Inguscio, Science \textbf{297},
2240 (2002).

\bibitem{gehm} M. E. Gehm, S. L. Hemmer, S. R. Granade, K. M. O'Hara, and J. E. Thomas, Phys. Rev. A {\bf68}, 011401(R) (2003).

\bibitem{heis} H. Heiselberg, Phys. Rev. A {\bf63}, 043606 (2001) 

\bibitem{fregoso} B. M. Fregoso and G. Baym, Phys. Rev. A {\bf 73}, 043616 (2006).

\bibitem{ODell_col} N. G. Parker, C. Ticknor, A. M. Martin, and D. H. J. O'Dell, Phys. Rev. A \textbf{79}, 013617 (2009).

\bibitem{bohi} O. Bohigas, A.M. Lane, and J. Martorell, Phys. Rep. {\bf51}, 267 (1979).

\bibitem{stringari2} E. Lipparini and S. Stringari, Phys. Rep. {\bf175}, 103 (1989).


\end{thebibliography}
\end{document}